\documentclass{article}

\usepackage{arxiv}

\usepackage[utf8]{inputenc} 
\usepackage[T1]{fontenc}    
\usepackage{hyperref}       
\usepackage{url}            
\usepackage{booktabs}       
\usepackage{amsfonts}       
\usepackage{nicefrac}       
\usepackage{microtype}      
\usepackage{lipsum}
\usepackage{graphicx}
\usepackage{caption}
\usepackage{subcaption}

\author{
  Anantha Natarajan \\
  Departments of Visualization \\
  Texas A\&M University\\
  College Station, Texas 77831 \\
  \texttt{ananthanatarajan@tamu.edu} \\
   \And
     Jiaqi Cui\\
  Departments of Visualization \\
  Texas A\&M University\\
  College Station, Texas 77831 \\
  \texttt{jiaqi96@tamu.edu } \\
   \And
  Ergun Akleman\thanks{website:people.tamu.edu/~ergun} \\
  Departments of Visualization \& \\Computer Science and Engineering\\
  Texas A\&M University\\
  College Station, Texas 77831 \\
  \texttt{ergun.akleman@gmail.com} \\
   \And
  Vinayak Krishnamurthy\\
  Departments of Mechanical Engineering \& \\Computer Science and Engineering\\
  Texas A\&M University\\
  College Station, Texas 77831 \\
  \texttt{ergun.akleman@gmail.com} \\
}

\title{Construction of Planar and Symmetric Truss Structures with Interlocking Edge Elements}

\date{}

\begin{document}

\maketitle

\thispagestyle{empty}

\begin{abstract}
In this paper, we present an algorithmic approach to design and construct planar truss structures based on symmetric lattices using modular elements. The method of assembly is similar to Leonardo grids as they both rely on the property of interlocking. In theory, our modular elements can be assembled by the same type of binary operations. Our modular elements embody the principle of geometric interlocking, a principle recently introduced in literature that allows for pieces of an assembly to be interlocked in a way that they can neither be assembled nor disassembled unless the pieces are subjected to deformation or breakage. We demonstrate that breaking the pieces can indeed facilitate the effective assembly of these pieces through the use of a simple key-in-hole concept. As a result, these modular elements can be assembled together to form an interlocking structure, in which the locking pieces apply the force necessary to hold the entire assembly together.


\end{abstract}
\begin{figure}[htpb]
    \centering
        \begin{subfigure}[t]{0.24\textwidth}
        \includegraphics[width=1.0\textwidth]{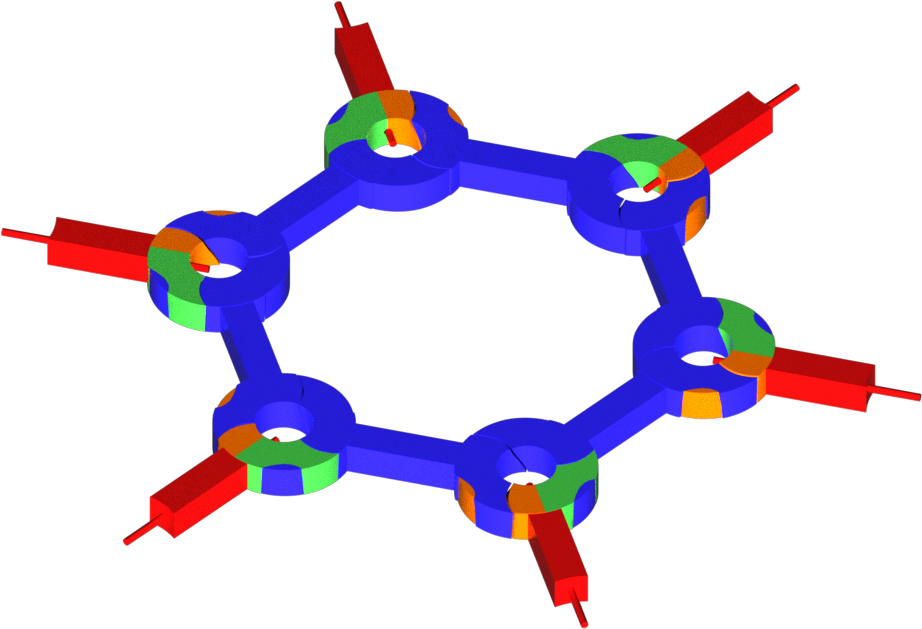}
        \caption{\it}
        \label{fig:Hexagonal/4_0}
    \end{subfigure}
    \hfill
            \begin{subfigure}[t]{0.24\textwidth}
        \includegraphics[width=1.0\textwidth]{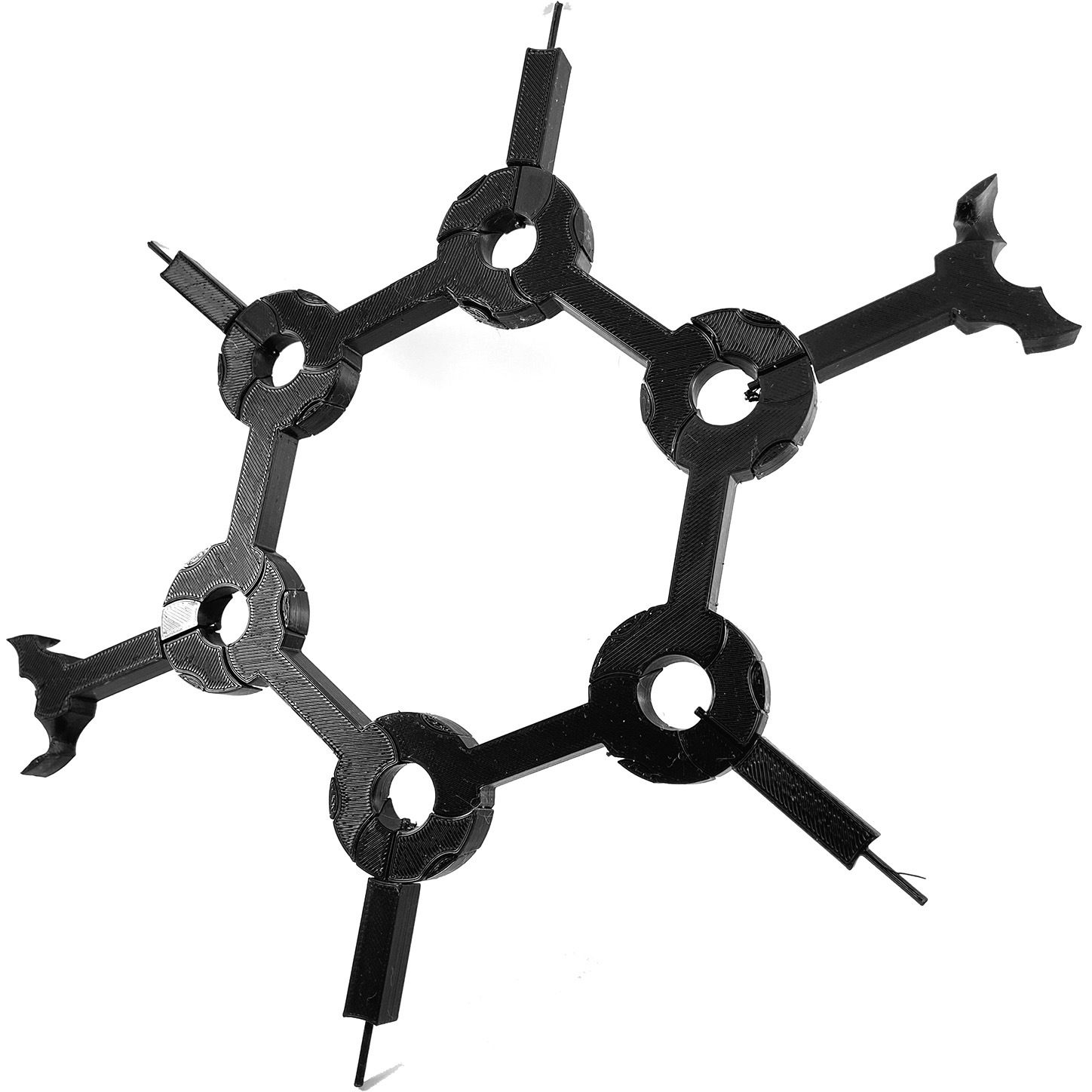}
        \caption{\it}
        \label{fig:Hexagonal/4_1}
    \end{subfigure}
    \hfill
            \begin{subfigure}[t]{0.24\textwidth}
        \includegraphics[width=1.0\textwidth]{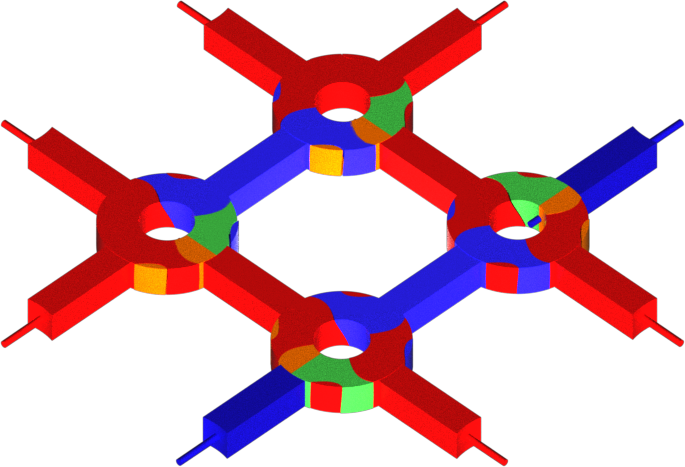}
        \caption{\it}
        \label{fig:Hexagonal/5_0}
    \end{subfigure}
    \hfill
            \begin{subfigure}[t]{0.24\textwidth}
        \includegraphics[width=1.0\textwidth]{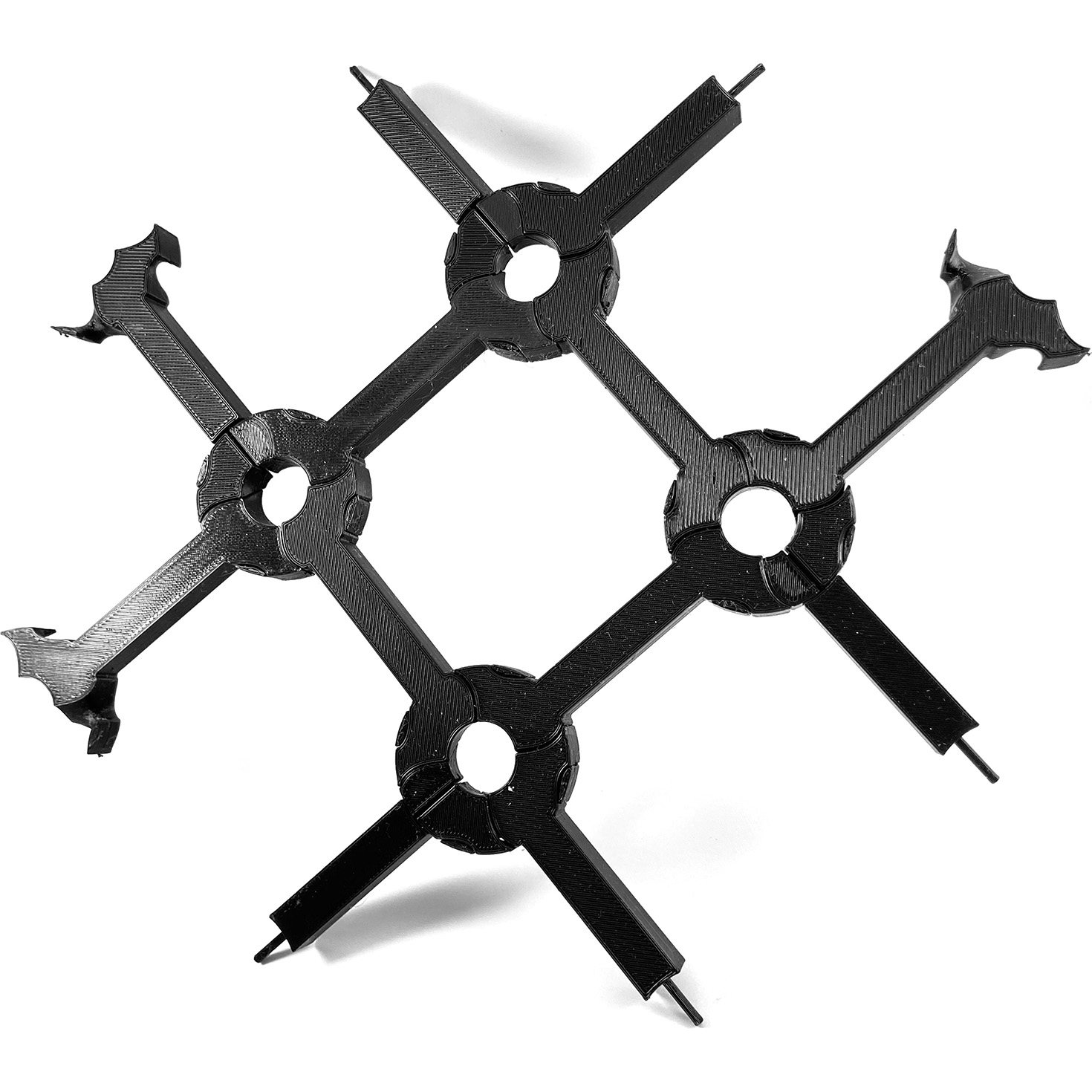}
        \caption{\it}
        \label{fig:Hexagonal/5_0}
    \end{subfigure}
    \hfill
\caption{\it Design and Fabrication of modular elements to construct 2D frames: (a) Design of Hexagonal frame, (b) Physical Assembly after fabrication, (c) Design of Quadrilateral frame, (d) Physical Assembly after fabrication}
\label{fig:HexagonalAssembly}
\end{figure}

\section{Introduction and Motivation}

History is rich with examples of puzzle-like interlocking structures that have been studied under the concepts of stereotomy \cite{evans1995,fallacara2009,fernando2015}, Leonardo grids or Nexorades \cite{baverel2000,douthe2009}, and topological interlocking  \cite{estrin2011,dyskin2001}. One of the most remarkable examples of interlocking structures is the Abeille flat vault, which was designed by the French architect and engineer Joseph Abeille \cite{fleury2009,akleman2020}. Leonardo grids or Nexorades are structures that are constructed using notched rods that fit into the notches of adjacent rods resembling fabric weaves \cite{song2013,roelofs2007}. 
\begin{figure}[htpb]
    \centering
    \begin{subfigure}[t]{0.15\textwidth}
        \includegraphics[width=1.0\textwidth]{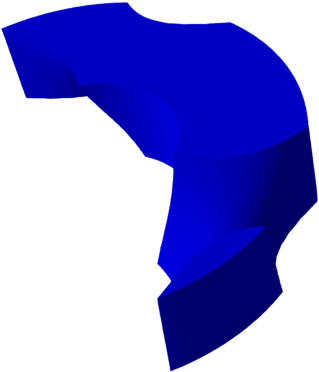}
        \caption{\it}
        \label{fig:Islamic_0/together}
    \end{subfigure}
        \hfill
    \begin{subfigure}[t]{0.15\textwidth}
        \includegraphics[width=1.0\textwidth]{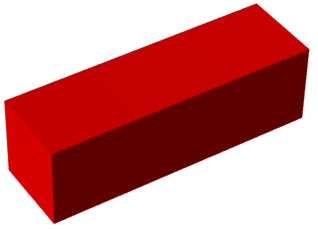}
        \caption{\it}
        \label{fig:Islamic_0/together}
    \end{subfigure}
        \hfill
    \begin{subfigure}[t]{0.3\textwidth}
        \includegraphics[width=1.00\textwidth]{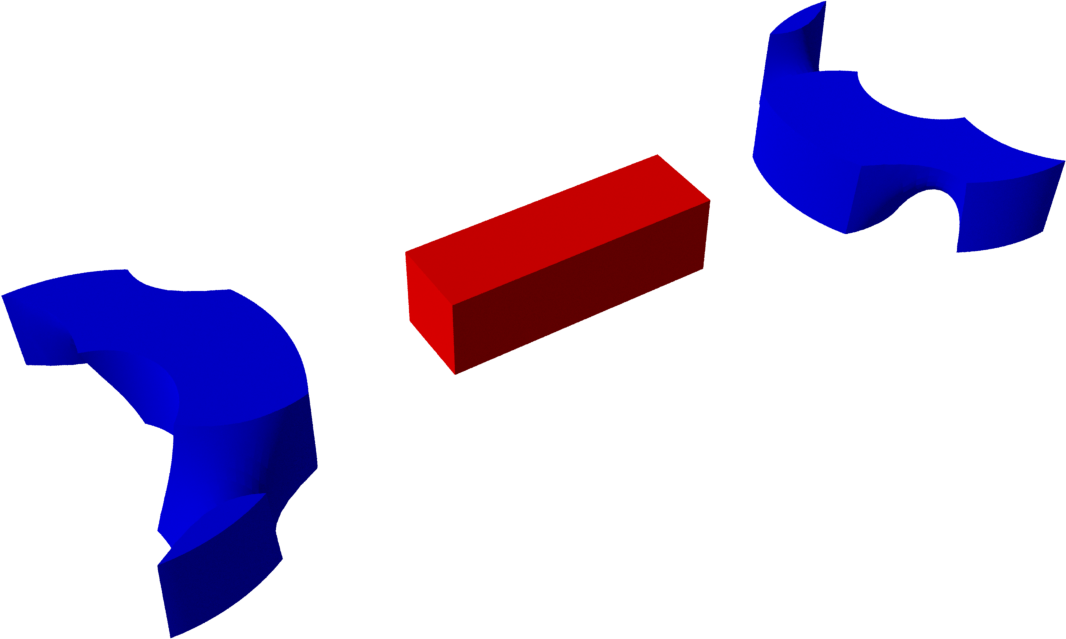}
        \caption{\it}
        \label{fig:Islamic/pattern}
    \end{subfigure}
    \begin{subfigure}[t]{0.3\textwidth}
        \includegraphics[width=1.00\textwidth]{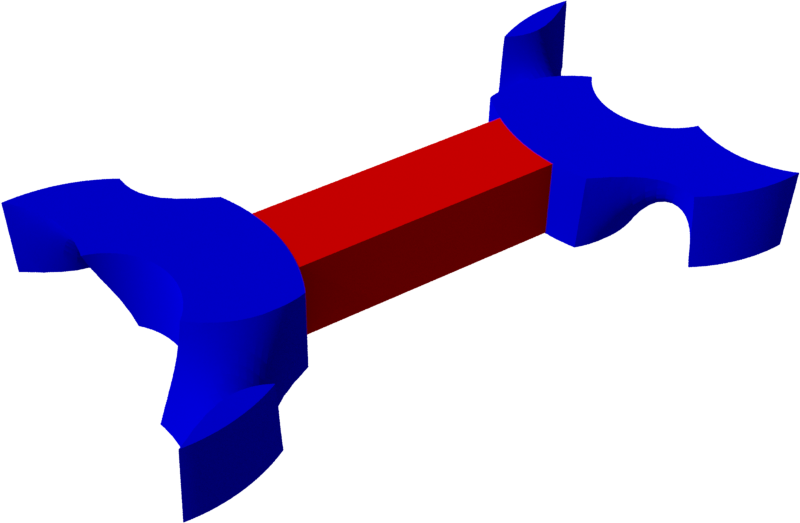}
        \caption{\it}
        \label{fig:Islamic/pattern_curved}
    \end{subfigure}
    \hfill
       
\caption{\it (a) Connector, (b) Truss Edge, (c) Creation of Edge Element, (d) Basic Edge Element  }
\label{fig:EdgeElementCreation}
\end{figure}

\begin{figure}[htpb]
    \centering
    \begin{subfigure}[t]{0.38\textwidth}
        \includegraphics[width=0.45\textwidth]{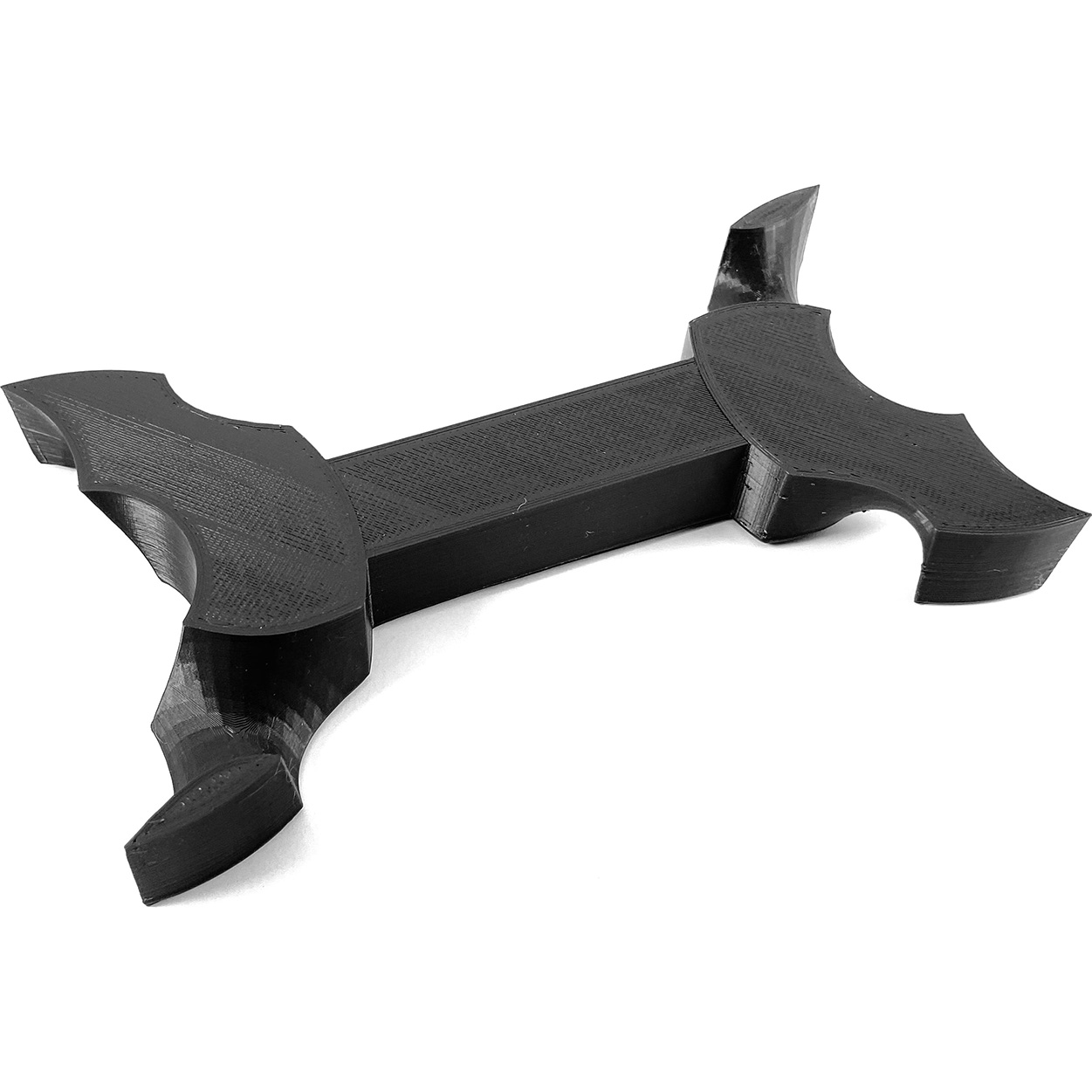}
        \includegraphics[width=0.45\textwidth]{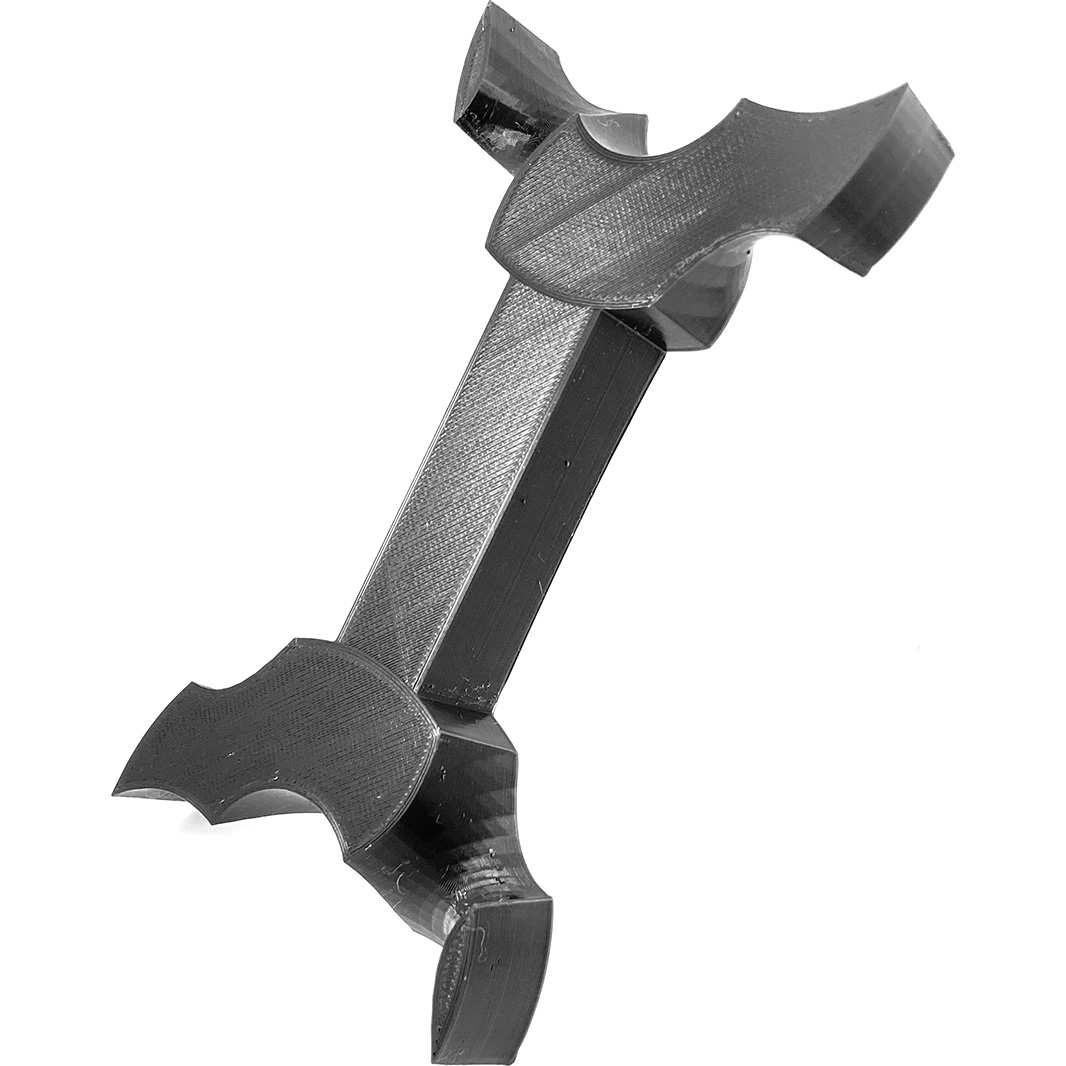}
        \caption{\it }
        \label{fig:Hexagonal/0_0}
    \end{subfigure}
    \hfill
        \hfill
    \begin{subfigure}[t]{0.19\textwidth}
        \includegraphics[width=1.0\textwidth]{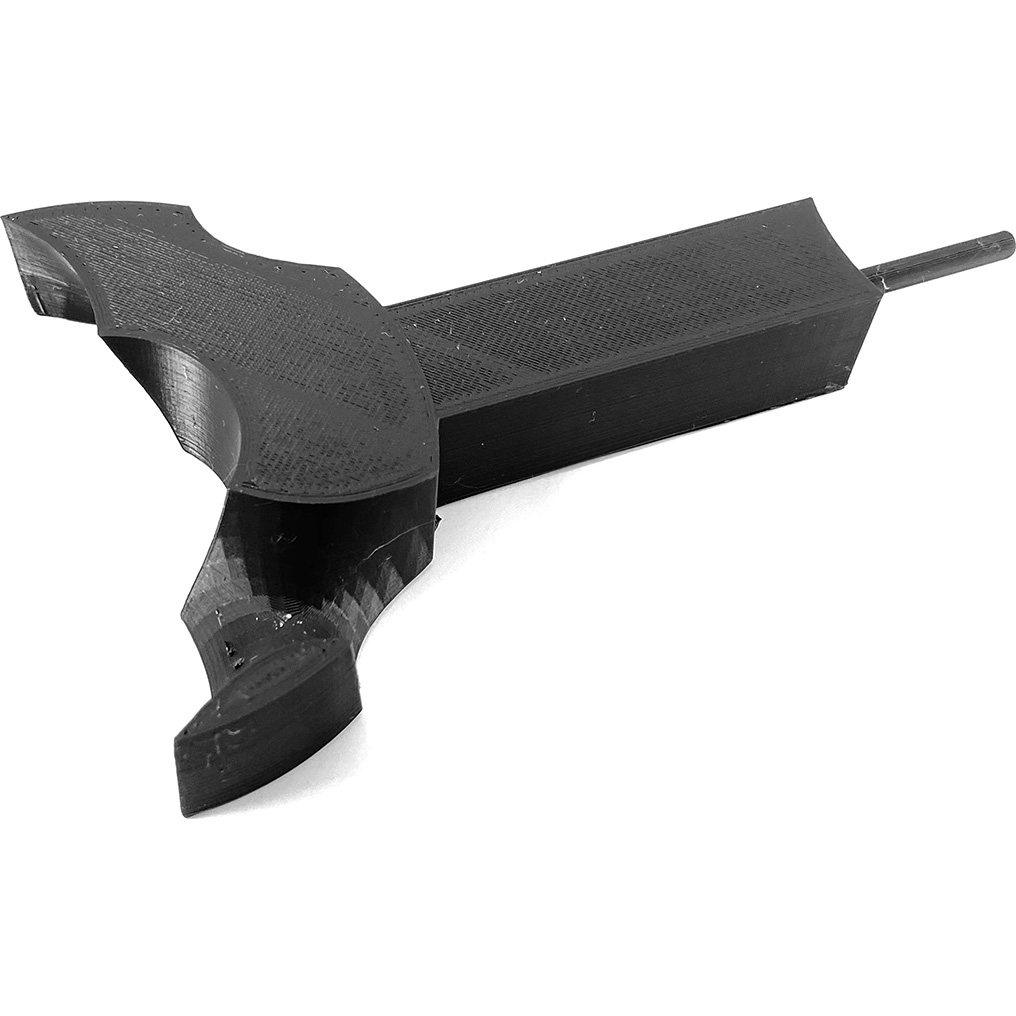}
        \caption{\it }
        \label{fig:Hexagonal/1_0}
    \end{subfigure}
    \hfill
        \begin{subfigure}[t]{0.19\textwidth}
        \includegraphics[width=1.0\textwidth]{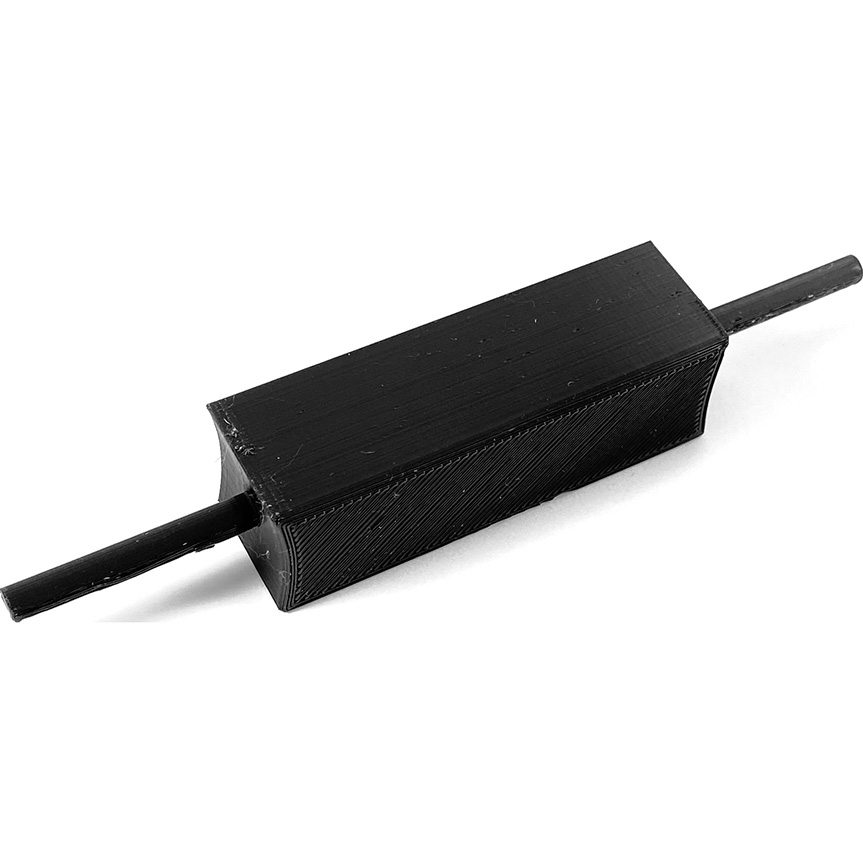}
        \caption{\it}
        \label{fig:Hexagonal/2_0}
    \end{subfigure}
    \hfill
        \begin{subfigure}[t]{0.19\textwidth}
        \includegraphics[width=1.0\textwidth]{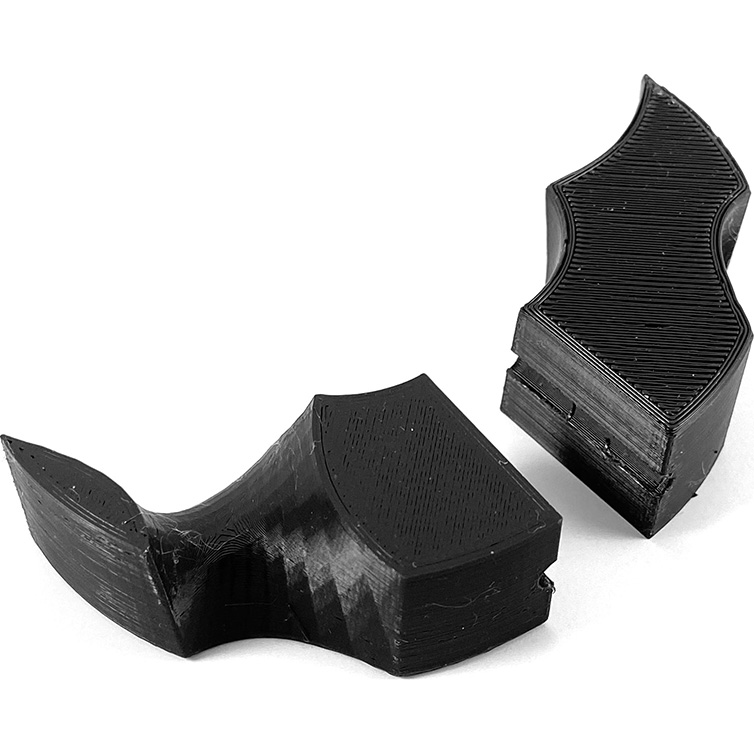}
        \caption{\it}
        \label{fig:Hexagonal/3_0}
    \end{subfigure}
    \hfill
\caption{\it Three types of interlocking Edge Elements and key pieces to construct 2D hexagonal frames: (a) Two views of a basic Edge Element, (b) Edge Element where one end is replaced by key, (c) Edge element with two keys, (d) Split Interlocking Connector that can locked by a key.}
\label{fig:HexagonalPieces}
\end{figure}

Symmetric patterns, such as Islamic ornaments, Ogee shapes, or Celcic knots, have been abundantly utilized as artistic decorations in architectural and sculptural design \cite{du2008symmetry,conway2016symmetries}. The symmetry in these patterns essentially allows for repetitions of geometric elements, thereby providing a sense of rhythm and aesthetic. 

\begin{figure}[htpb]
    \centering
        \begin{subfigure}[t]{0.32\textwidth}
        \includegraphics[width=1.0\textwidth]{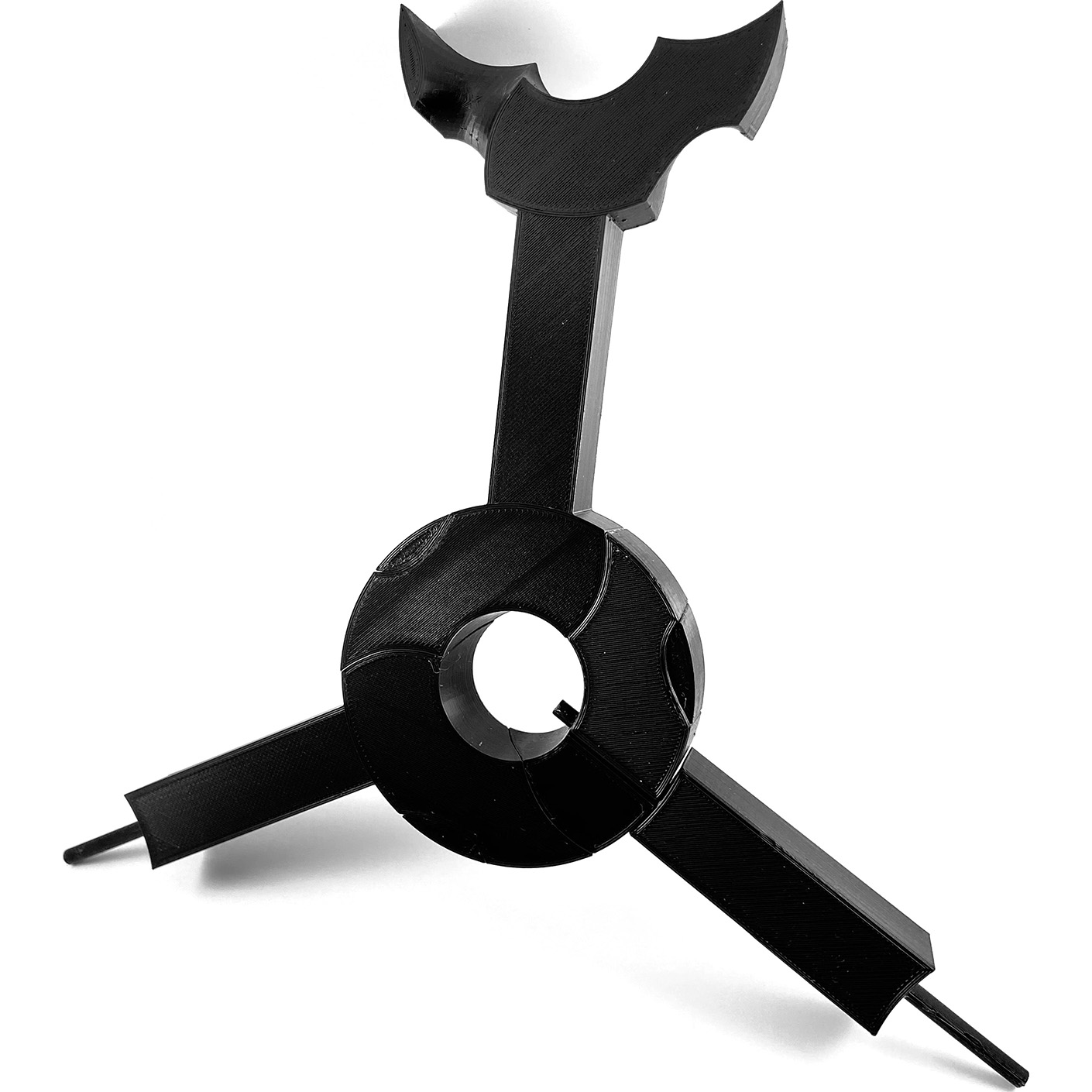}
        \caption{\it}
        \label{fig:Hexagonal/4_0}
    \end{subfigure}
    \hfill
            \begin{subfigure}[t]{0.32\textwidth}
        \includegraphics[width=1.0\textwidth]{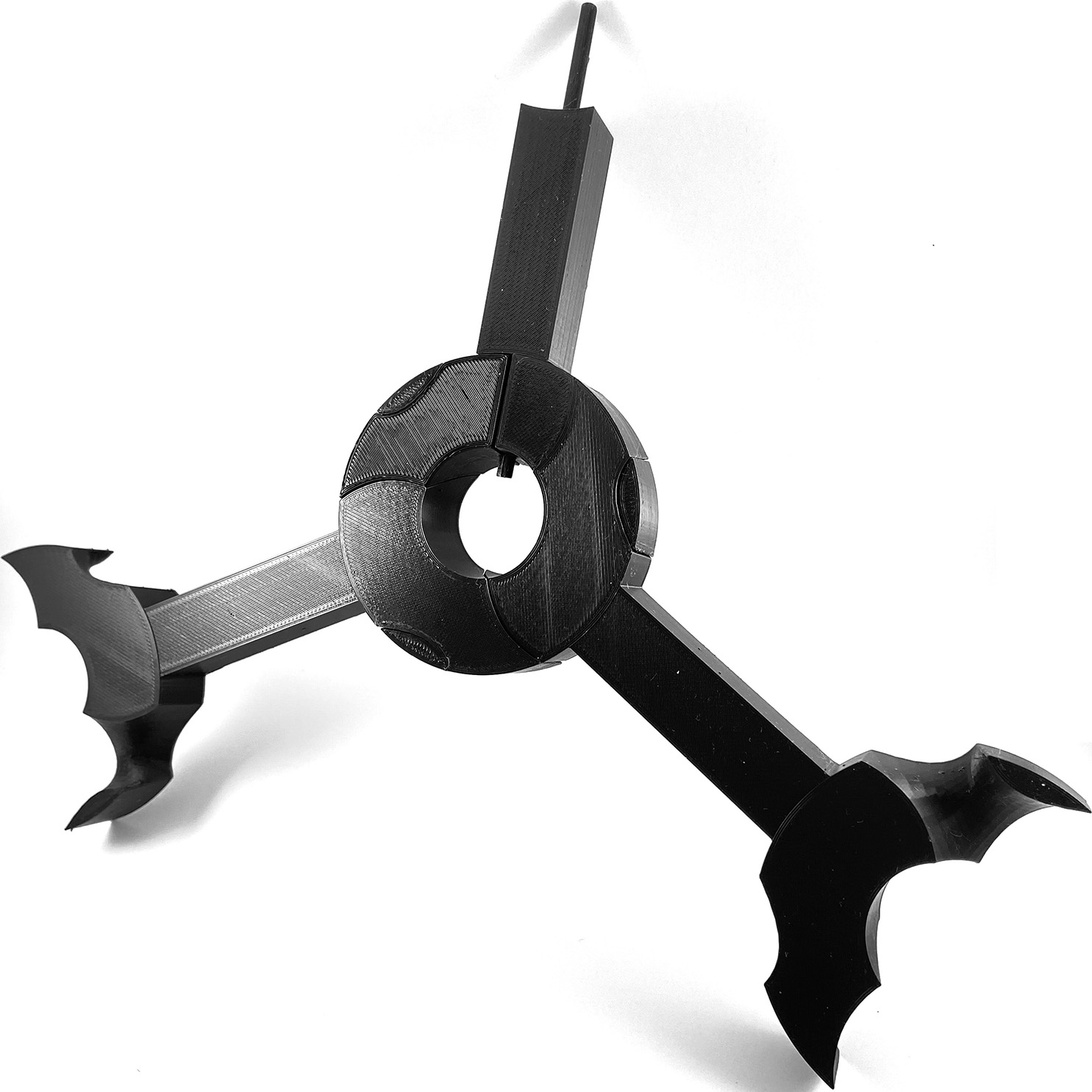}
        \caption{\it}
        \label{fig:Hexagonal/4_1}
    \end{subfigure}
    \hfill
            \begin{subfigure}[t]{0.32\textwidth}
        \includegraphics[width=1.0\textwidth]{images/hexagonal/5_0}
        \caption{\it}
        \label{fig:Hexagonal/5_0}
    \end{subfigure}
    \hfill
\caption{\it Assembly of Edge Elements shown in Figure~\ref{fig:HexagonalPieces} to construct of 2D hexagonal frames: (a) Assembly of a three sided corner that use all types of Edge Elements, (b) Assembly of a three sided corner that use two basic Edge Element and one Edge Element with two keys, (c) Assembly of one hexagon. This structure can indefinitely be extended to form a hexagonal grid}
\label{fig:HexagonalAssembly}
\end{figure}

\begin{figure}[htpb]
    \centering
    \begin{subfigure}[t]{0.15\textwidth}
        \includegraphics[width=1.0\textwidth]{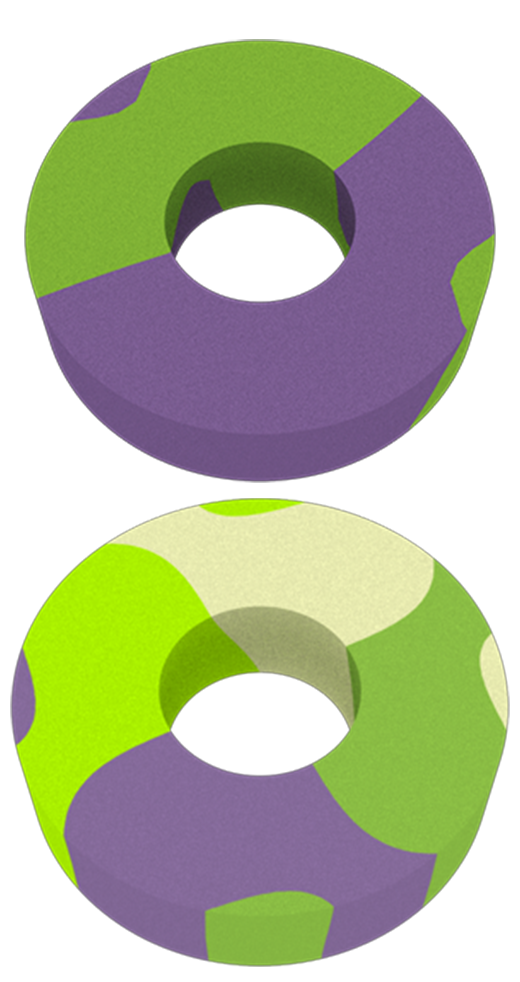}
        \caption{\it}
        \label{fig:Islamic_0/together}
    \end{subfigure}
        \hfill
    \begin{subfigure}[t]{0.40\textwidth}
        \includegraphics[width=1.00\textwidth]{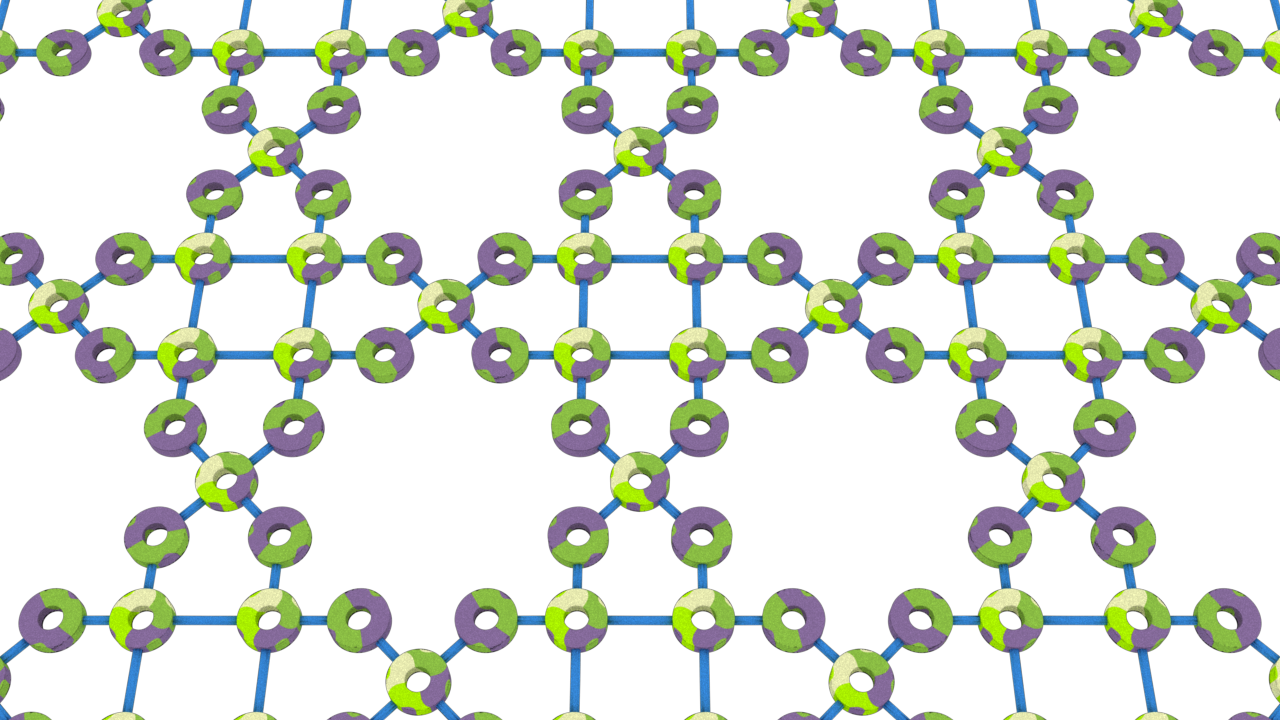}
        \caption{\it}
        \label{fig:Islamic/pattern}
    \end{subfigure}
    \begin{subfigure}[t]{0.40\textwidth}
        \includegraphics[width=1.00\textwidth]{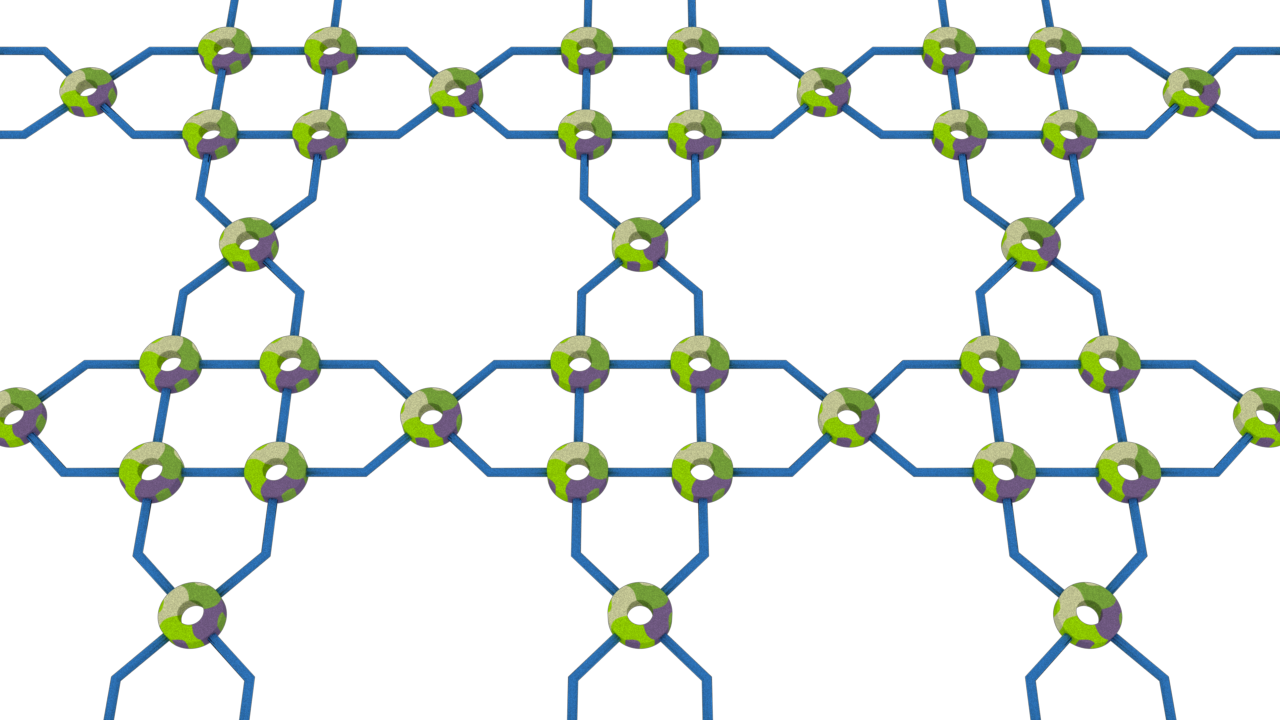}
        \caption{\it}
        \label{fig:Islamic/pattern_curved}
    \end{subfigure}
    \hfill
       
\caption{\it An Islamic pattern generated by our methodology: (a) Two types of toroidal decomposition, (b) Pattern created using straight edges and both types of toroidal decomposition, (c) The same pattern created with bent edges and one type of toroidal decomposition}
\label{fig:Islamic Pattern Example}
\end{figure}

In this work, we have developed a new concept to produce symmetric 2D patterns that are constructed as interlocked truss structures. Our concept is based on designing modular \textit{Edge Elements} that can be assembled to create symmetric planar trusses and frames. These modular Edge Elements support each other by virtue of their geometry without the need for any adhesive or binding materials. Our underlying principle is to decompose the joints of a given truss/frame as a set of interlocking shapes. These interlocking shapes which we call \textit{Connectors}, in combination with the edges of the truss, result in \textit{Edge Elements} (Figure~\ref{fig:EdgeElementCreation}). These Edge Elements can be connected with each other to physically build the truss (Figure~\ref{fig:HexagonalAssembly}). While we demonstrate these Edge Elements through examples of trusses based on uniform and regular planar tessellations, our approach as such can be easily expanded for complex 3D trusses to create free form and free spanning grid shell structures and architectural designs such as Islamic patterns (See Figure~\ref{fig:Islamic Pattern Example}).

\section{Previous Work}
Our Edge Elements are inspired from our recent methodology for designing interlocking structures using higher-dimensional Voronoi sites  \cite{subramanian2019,akleman2020}. Allowing any type of shape to serve as Voronoi sites provides \textbf{a simple and systematic design methodology} to construct a large variety of interlocking structures for any volumetric domain in 3-space. Based on this point of view, the key parameters are essentially the topological and geometric properties of Voronoi sites and their overall arrangements that are usually be obtained by symmetry transformations (rotation, translation, and mirror operations). The types of shapes that serve as Voronoi sites and their transformations uniquely determine the properties of how the space is partitioned. For instance, let us consider certain special types of partitions such as  Delaunay's Stereohedra \cite{delaunay1961,schmitt2016}, the Delaunay lofts, generalized Abeille tiles, and bi-axial woven tiles \cite{subramanian2019,akleman2020,krishnamurthy2020}. 

For Stereohedra, the shapes of Voronoi sites are points, 3D $L_2$ norm is used for distance computation, underlying space is 3D, and any symmetry operation in 3D is allowed \cite{delaunay1961,schmitt2016}. Based on these properties, we conclude that Stereohedra can theoretically represent every convex space-filling polyhedra in 3D. Since the points are used as Voronoi sites, and $L_2$ norm is used, the faces must be planar, and edges must be straight in the resulting Voronoi decomposition of the 3D space.

For Delaunay lofts, on the other hand, the shapes of Voronoi sites are curves that are given in the form of $x=f(z)$ and $y=g(z)$, for every planar layer $z=c$ where $c$ is a real constant, a 2D $L_2$ norm is used to compute distance, underlying space is 2.5 or 3D, and only 17 wallpaper symmetries are allowed in every layer $z=c$ \cite{subramanian2019}. Based on these properties, we conclude that Delaunay lofts (1) consist of stacked layers of planar convex polygons with straight edges, and (2) in each layer, there can be only one convex polygon. In Delaunay lofts, the number of sides of the stacked convex polygons can change from one layer to another. In conclusion, the faces of the Delaunay lofts are ruled surfaces since they consist of sweeping lines. Edges of the faces can be curved. For generalized Abeille tiles, Voronoi sites can be ruled surfaces or tree-structures, which can significantly extend design space \cite{akleman2020}. However, they do not provide geometric interlocking properties.  

If the shapes of Voronoi sites are curve segments obtained by decomposing planar periodic curves that are closed under symmetries of bi-axial weaving patterns, the result becomes geometrically interlocking assemblies such as Bi-axial woven tiles \cite{krishnamurthy2020}. In this paper, we use this particular type of Voronoi decomposition to create our Connectors. 
Using these Connectors as basic building blocks, we construct all types of fundamental modular blocks that we call Edge Elements (see Figure~\ref{fig:HexagonalPieces}). 




\section{Methodology} 

Our process starts with a planar graph that represents the truss/frame topology. In this paper, we demonstrate the basic idea through the square and regular hexagonal grids as our truss topology. In our process, all edges of the grid are replaced by square prisms, and all vertices are replaced by a 1-holed toroidal surface with a square cross-section. This toroidal surface is constructed by rotating a square about an axis parallel to a vertical side of the square for a given radius. The toroidal surface encloses a volumetric domain that serves as the interlocking region. Our goal is to decompose this volumetric domain into geometrically interlocked pieces. There exists a variety of methods to obtain topologically, or geometrically interlocking pieces for a given domain, such as Delaunay lofts \cite{subramanian2019}, generalized Abeille tiles \cite{akleman2020}, or bi-axial woven tiles \cite{krishnamurthy2020}. In this work, we decompose the domain using an extension of the method that is used to design bi-axial woven tiles. Our process decomposes the toroidal volumetric domain into geometrically interlocking woven tiles which we call Connectors and use these Connectors to create Edge Elements, and it consists of five steps.

\begin{figure}[htpb]
    \begin{subfigure}[t]{0.20\textwidth}
        \includegraphics[width=1.0\textwidth]{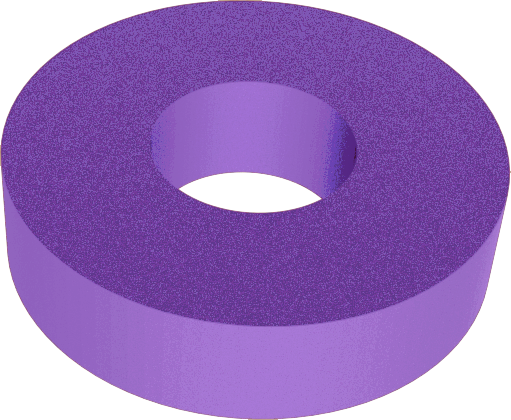}
        \caption{\it}
        \label{fig:images/process/step0}
    \end{subfigure}
    \hfill
 \begin{subfigure}[t]{0.20\textwidth}
        \includegraphics[width=1.0\textwidth]{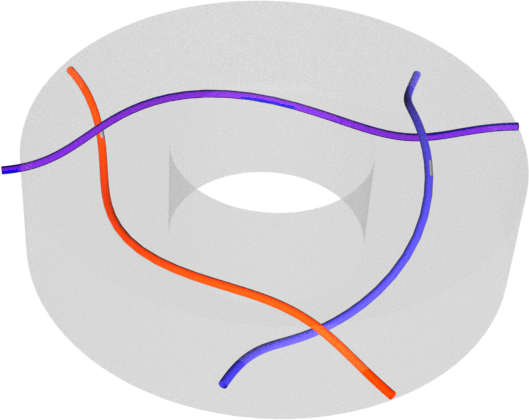}
        \caption{\it}
        \label{fig:images/process/step1}
    \end{subfigure}
    \hfill
     \begin{subfigure}[t]{0.20\textwidth}
        \includegraphics[width=1.0\textwidth]{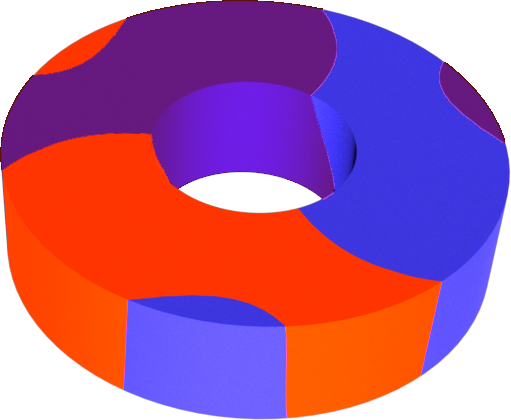}
        \caption{\it}
        \label{fig:images/process/step2}
    \end{subfigure}
    \hfill   
    \begin{subfigure}[t]{0.24\textwidth}
        \includegraphics[width=1.0\textwidth]{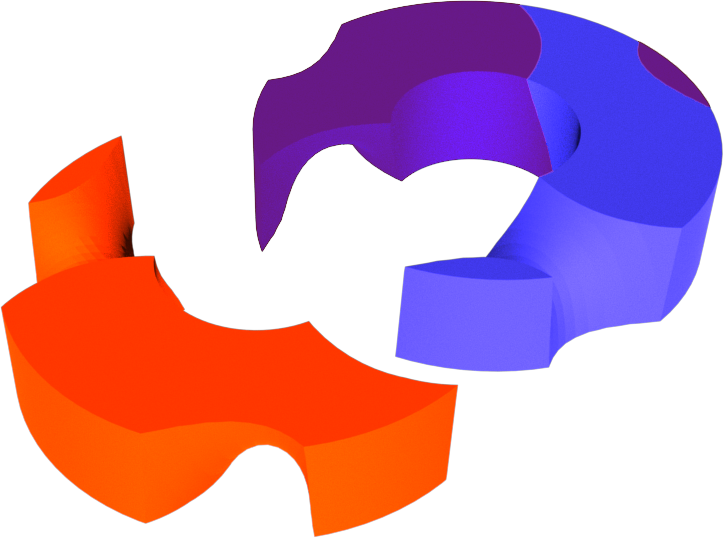}
        \caption{\it}
        \label{fig:images/process/step3}
    \end{subfigure}
    \hfill 
        \begin{subfigure}[t]{0.32\textwidth}
        \includegraphics[width=1.0\textwidth]{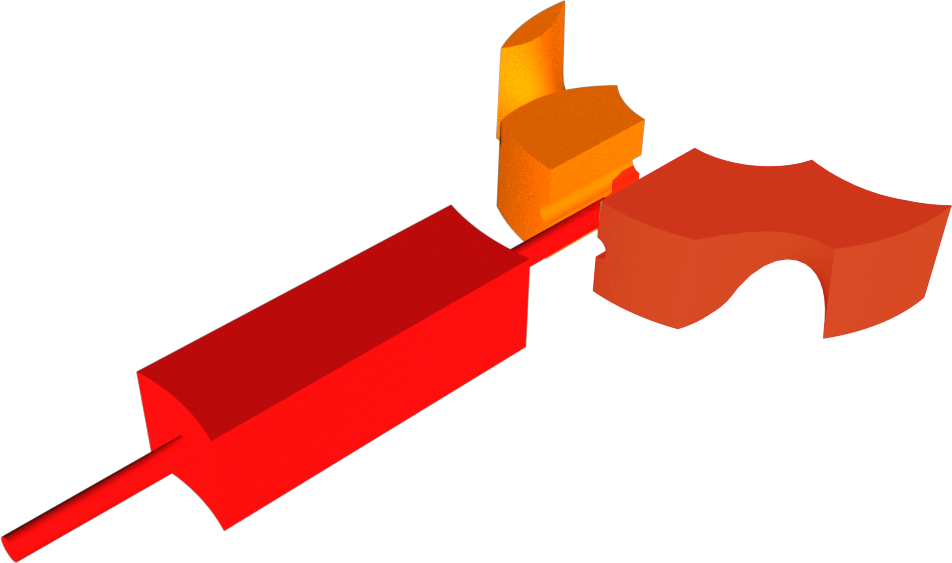}
        \caption{\it}
        \label{fig:images/process/step4}
    \end{subfigure}
    \hfill 
        \begin{subfigure}[t]{0.32\textwidth}
        \includegraphics[width=1.0\textwidth]{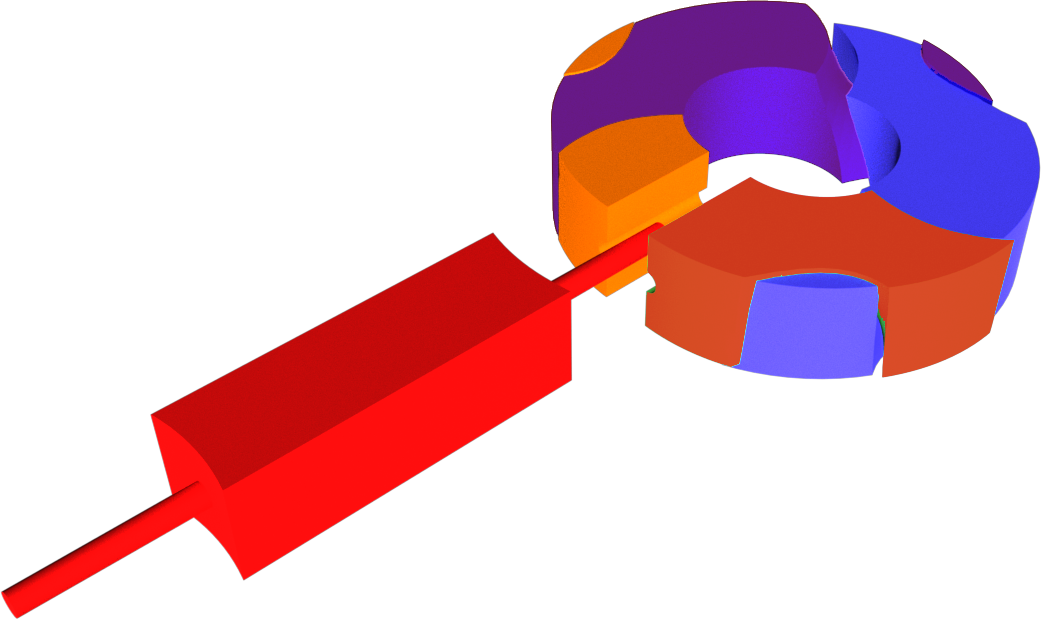}
        \caption{\it}
        \label{fig:images/process/step5}
    \end{subfigure}
    \hfill 
        \begin{subfigure}[t]{0.32\textwidth}
        \includegraphics[width=1.0\textwidth]{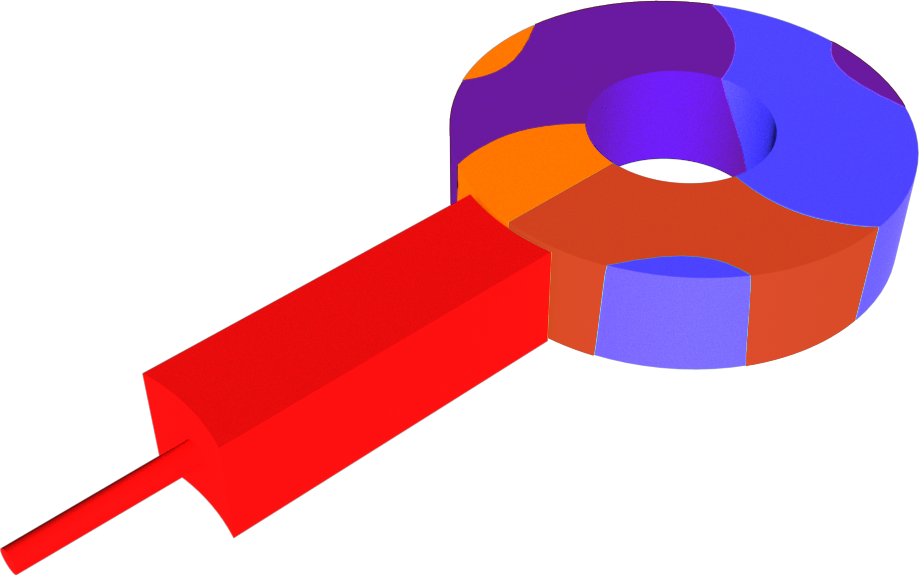}
        \caption{\it}
        \label{fig:images/process/step6}
    \end{subfigure}
    \hfill 
\caption{\it Visuals that demonstrate the steps of the process: (a) The initial 1-holed toroidal surface with a square cross section, (b) Sinusoidal curves as Voronoi Sites, (c) Voronoi decomposition of the toroidal domain using Sinusoidal curves as Voronoi Sites, (d) One of the Connectors (red one) must be further decomposed, (e)The red Connector is further decomposed to be used as locking pieces, (f) Assembly process, (g) Complete assembly of Connectors}
\label{fig:processsteps}
\end{figure}

\begin{figure}[htpb]
    \centering
    \begin{subfigure}[t]{0.22\textwidth}
        \includegraphics[width=1.0\textwidth]{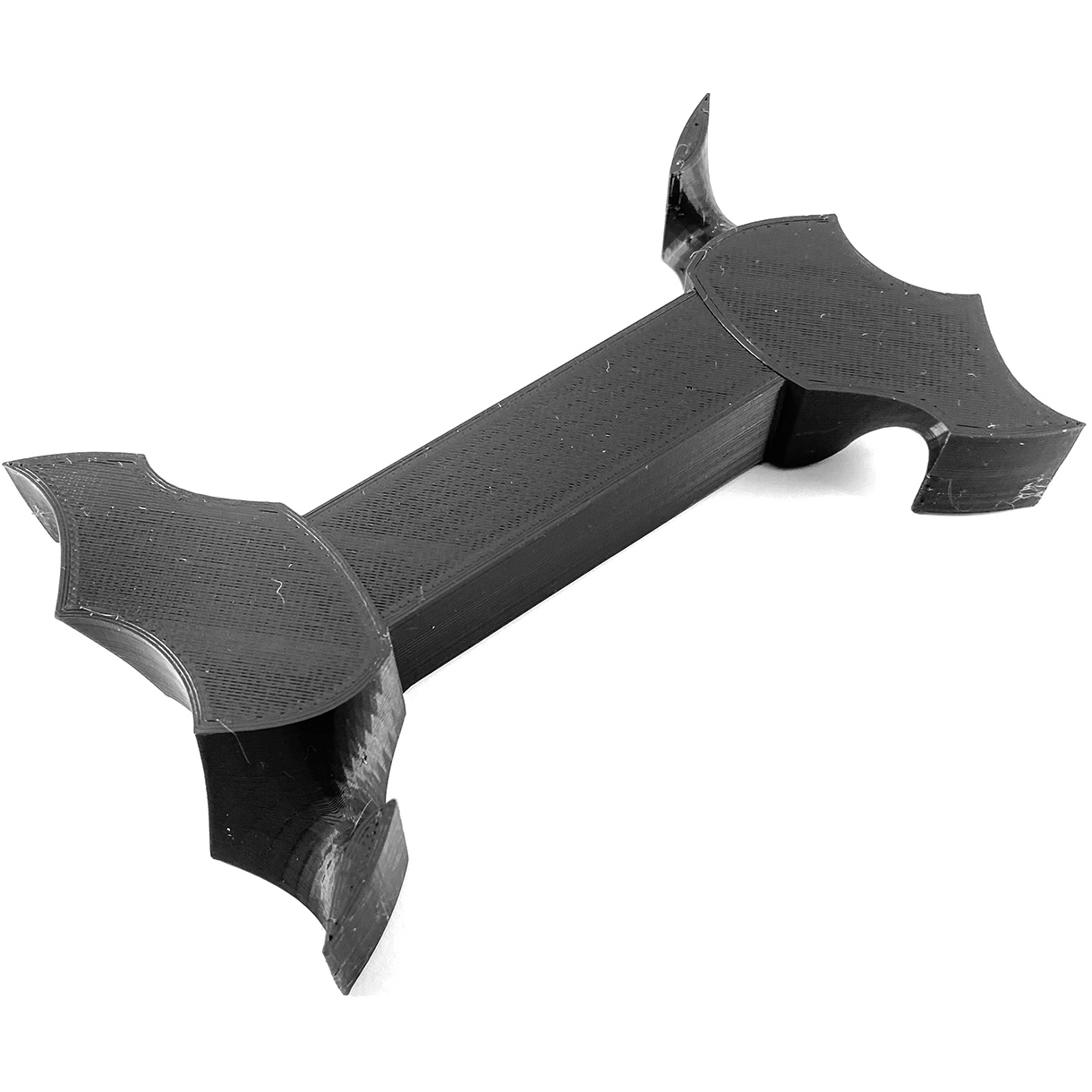}
        \caption{\it}
        \label{fig:quadrilateral/0_0}
    \end{subfigure}
    \hfill
        \hfill
    \begin{subfigure}[t]{0.22\textwidth}
        \includegraphics[width=1.0\textwidth]{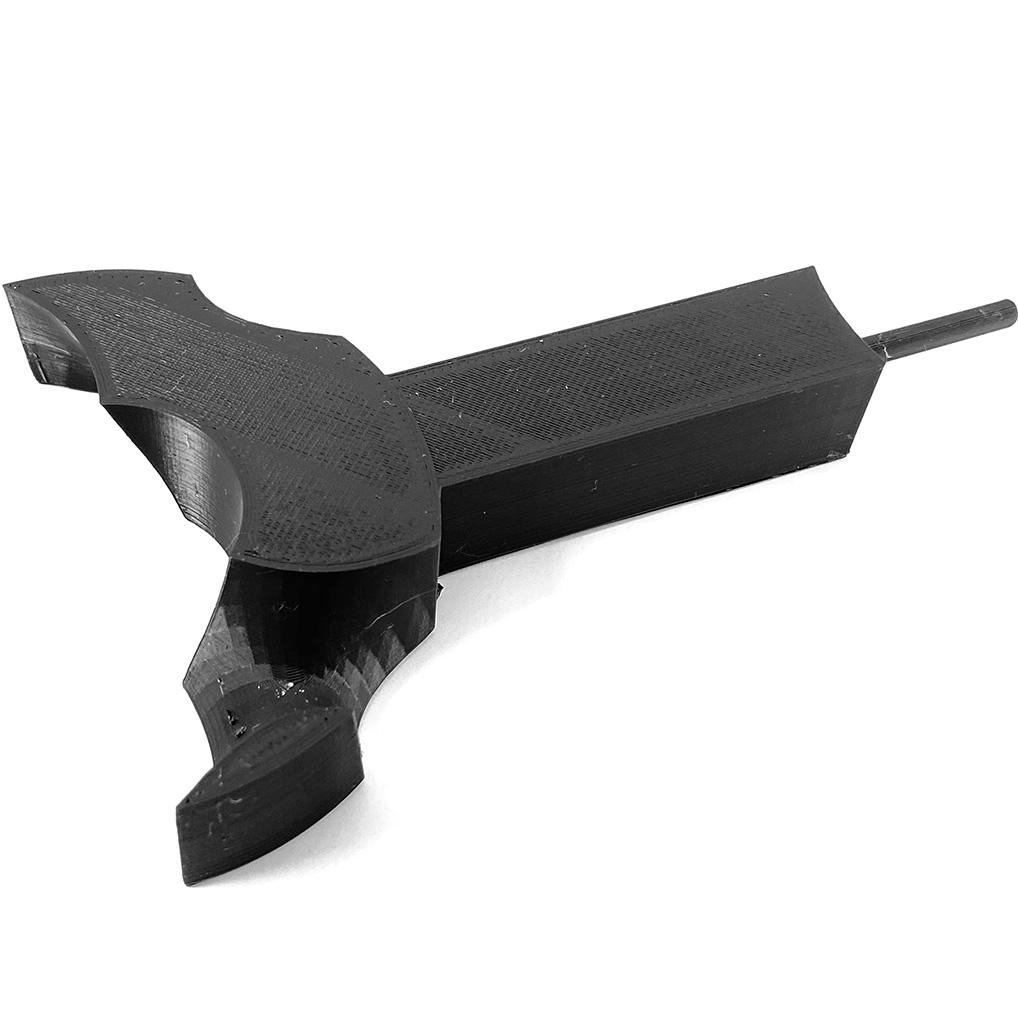}
        \caption{\it}
        \label{fig:quadrilateral/1_0}
    \end{subfigure}
    \hfill
        \begin{subfigure}[t]{0.22\textwidth}
        \includegraphics[width=1.0\textwidth]{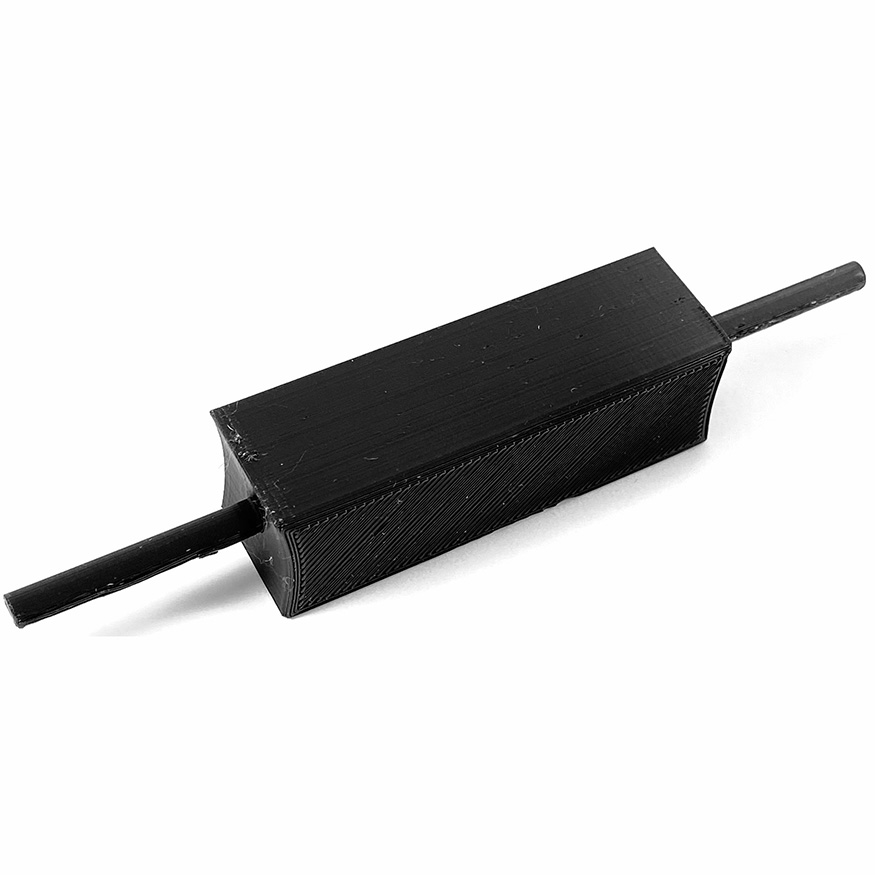}
        \caption{\it }
        \label{fig:quadrilateral/2_0}
    \end{subfigure}
    \hfill
        \begin{subfigure}[t]{0.22\textwidth}
        \includegraphics[width=1.0\textwidth]{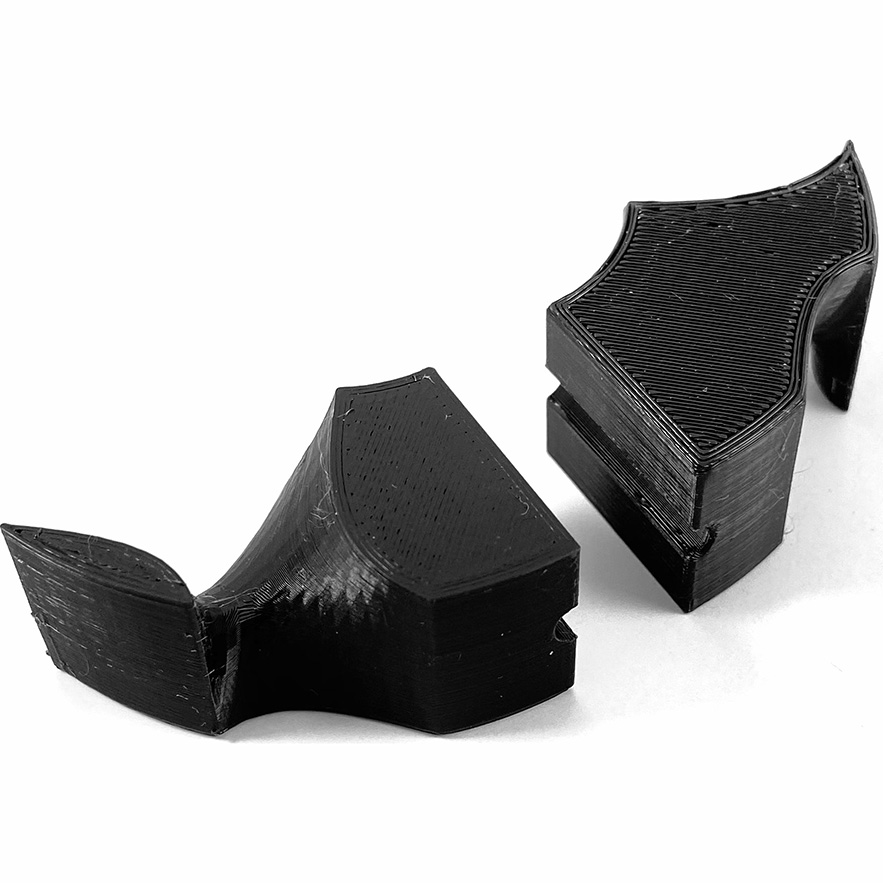}
        \caption{\it }
        \label{fig:quadrilateral/3_0}
    \end{subfigure}
    \hfill
        \begin{subfigure}[t]{0.22\textwidth}
        \includegraphics[width=1.0\textwidth]{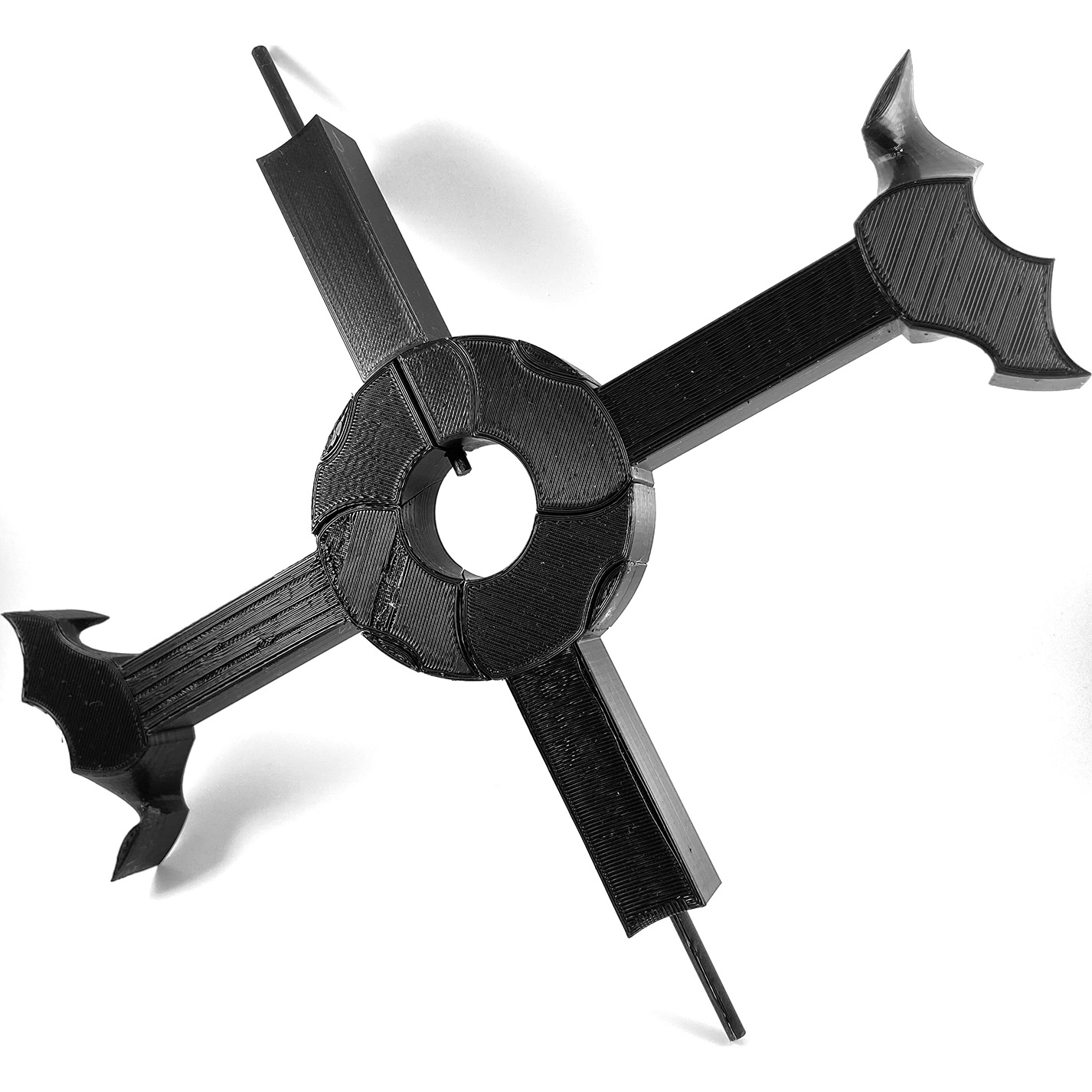}
        \caption{\it}
        \label{fig:quadrilateral/4_0}
    \end{subfigure}
    \hfill
            \begin{subfigure}[t]{0.22\textwidth}
        \includegraphics[width=1.0\textwidth]{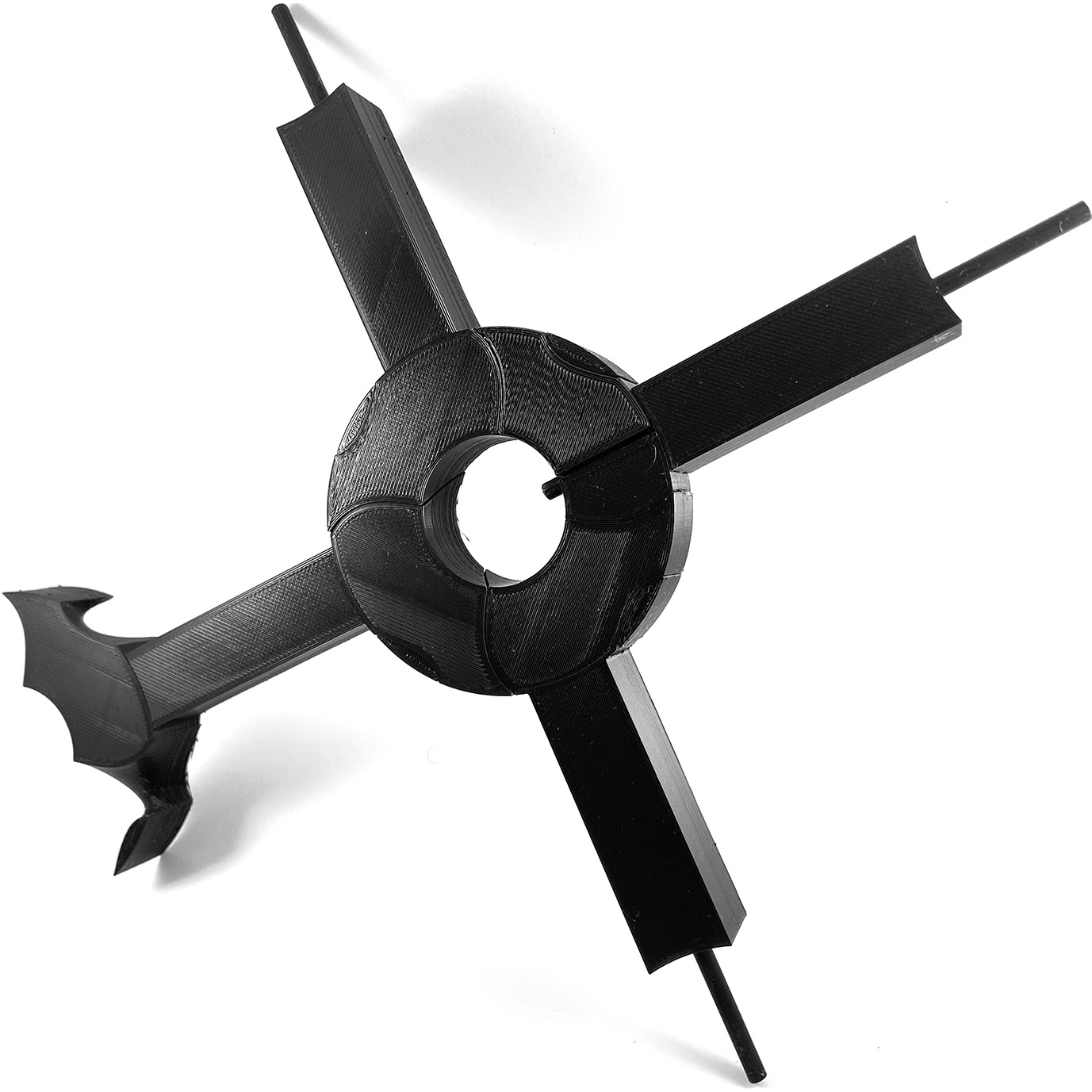}
        \caption{\it}
        \label{fig:quadrilateral/4_1}
    \end{subfigure}
    \hfill
            \begin{subfigure}[t]{0.22\textwidth}
        \includegraphics[width=1.0\textwidth]{images/quadrilateral/5_0}
        \caption{\it}
        \label{fig:quadrilateral/5_0}
    \end{subfigure}
    \hfill
\caption{\it Interlocking Edge Elements and key pieces to construct 2D quadrilateral frames: (a) A basic Edge Element for quadrilateral grid, (b) One sided locking Edge Element, (c) Two sided locking Edge Element, (d) Split Connector tiles, (e) An assembly of a four sided corner that use all the pieces, (f) Another assembly of a four sided corner that use all the pieces, (g) A complete assembly of a single quadrilateral}
\label{fig:quadrilateral}
\end{figure}

\begin{enumerate}
\item \textbf{Voronoi Site Generation:} The key step in our process is the generation of higher dimensional Voronoi sites. In this case, we use sinusoidal curves as the higher dimensional Voronoi sites to create geometrically interlocking pieces. The number of rotations and the rotation angles depend on the grid pattern to be constructed. For regular hexagonal grids, each vertex is shared by three hexagons. The angle between the edges that share the same vertex is $120$ degrees. The toroidal domain at this vertex (Figure~\ref{fig:images/process/step0}) should be decomposed into three identical geometrically interlocking pieces that can be attached to a polygon edge. This means that the initial sinusoidal curves must be rotated $120$ degrees about the toroidal domain’s axis of revolution. Performing this rotation $3$ times gives us three sinusoidal curves that are non-intersecting and non-planar (see Figure~\ref{fig:images/process/step1}). These curves are then deformed uniformly from the axis of revolution to match the horizontal curvature of the torus. For a vertex with a different valency, the weave pattern generated can be adjusted to create geometrically interlocking tiles by changing the length, amplitude, amount of deformation and number of the curves.
\item \textbf{Voronoi Decomposition to Obtain Connectors:} Voronoi decomposition is generally described for points as Voronoi sites. In order to compute Voronoi decomposition for our sinusoidal curves, we draw from previous works \cite{subramanian2019,akleman2020,krishnamurthy2020} and sample points on these curves. We subsequently label all the points belonging to the same curve into a cluster. As a  result, we obtain a group of co-labeled Voronoi sites for each curve. We then obtain Voronoi decomposition of the domain using these points. This process decomposes the domain into convex polyhedral regions. Taking the union of the Voronoi cells corresponding to co-labeled sites (i.e., sites belonging to the same curve) results in the partitioning of the toroidal domain into three interlocking woven tiles (Connectors) in a hexagonal grid case. The shapes of these Connectors are identical because of the rotational symmetry of the curves (See Figure~\ref{fig:images/process/step1}). \textbf{Remark:} For constructing a square grid pattern, the domain is decomposed into $4$ identical Connectors, which means the sinusoidal curve is rotated $90$ degrees $4$ times to produce $4$ non intersecting, non-planar control curves. 

\item \textbf{Creation of Locking Pieces by Cutting one of the Connectors:}  The resulting structure is geometrically interlocking (shown in Figure~\ref{fig:images/process/step2}), and it cannot be assembled without cutting one of the Connectors. We therefore choose one Connector (See the red tile Figure~\ref{fig:images/process/step3}) and decompose it into three pieces.  In our implementation, decomposition is achieved by vertically splitting the Connector into two equal halves with a socket to house the locking peg  (See the decomposition in Figure~\ref{fig:images/process/step4}). The pair of half tiles serve as the locking pieces. The peg that is vertically and horizontally centered is shared between the halves and plays the role of a key that holds the assembly(See Figures~\ref{fig:images/process/step5} and ~\ref{fig:images/process/step6}).

\item \textbf{Creation of Edge Elements} The interlocking Connectors are joined together by a square tube to obtain Edge Elements. The side of the square tube is equal to the side of the cross-section of the toroidal domain. Apart from the locking pieces mentioned in the previous step, there are three types of Edge Elements needed to create any grid pattern. The basic Edge Element type is constructed by attaching a Connector (the original tiles obtained by Voronoi decomposition) to both ends of a square tube. In addition, there are two locking Edge Element types- (1) one-sided locking Edge Element - square tube with the interlocking Connector on one end and a locking peg on the other (2) two-sided locking Edge Element - square tube with a locking peg on both ends.

\item \textbf{Printing and Shape Assembly} In order to construct physical assemblies of truss structures, we printed multiple copies of each type of Edge Element and assembled them to create structures as shown in Figures~\ref{fig:Hexagonal/4_0} and ~\ref{fig:Hexagonal/4_1}. This assembly is repeated multiple times as needed and put together to create the grid pattern as shown in Figure~\ref{fig:Hexagonal/5_0}. We have created two types of grids: regular hexagon and square as shown in Figures~\ref{fig:Hexagonal/5_0}  and  \ref{fig:quadrilateral/5_0} respectively. In order to create square grid, we designed the Connectors using four sinusoidal curves and printed the Edge Elements similarly (Figure 6a, 6b, 6c, 6d). Resulting vertex assemblies are shown in \ref{fig:quadrilateral/4_0} and \ref{fig:quadrilateral/4_1} and grid assembly is shown in \ref{fig:quadrilateral/5_0}.


\end{enumerate}

\section{Discussion}
Our preliminary observations show that locking the connectors with keys results in a stable and sturdy assembly. The main advantage of this method is that our Edge Elements can be duplicated economically by mass production such as casting. The modular property of our Edge Elements makes transportation and construction simpler. Our Edge Elements makes the creation of symmetric patterns in architectural and sculptural structures efficient. In a different perspective, they can also be used as construction toys that can help build cognitive ability in kids. However, this work is still preliminary, and there is a lot more that can be done to extend this work. We list some of these extensions that can be practically useful below.

\begin{itemize}
\item \textbf{Trusses based on Uniform Regular Tessellations:} Our method is not limited to regular mesh structures such as square or regular hexagonal grids. The idea in this paper can be generalized to any grid in which all the vertices are isometric with each other. The simplest of such grids are rectangular grids. This space of grids is very rich. However, since the angles between consecutive edges around the vertex may not necessarily be the same, there is a need for additional care in organizing sinusoidal curves. The main advantage of this type of grids, the elements can also be mass-produced since they can consist of a very limited set of Edge Elements.  
\item \textbf{Trusses based on Semi-regular and Irregular Tessellations:} The method can directly be extended for trusses based on general tessellations of the plane. Here, we note that the regularity or uniformity (or the lack thereof) of the tessellation would dictate whether we can mass produce a finite set of repeating Edge Elements or if we each Edge Element is completely different from all others (as would happen for a completely arbitrary tessellation of the plane).
\item \textbf{Regular Polyhedral Trusses:} We observe that the repetitive characteristic of our Edge Elements for regular shapes can be effectively utilized for regular polyhedra, i.e. platonic solids. Since all vertices in Platonic are isomorphic to each other, the Edge Elements can be mass-produced. The sharp angle that exists between edges can be handled by changing the cross-section of the toroidal domain. If a right tangential trapezoid is used instead of the square as the cross-section, the slanted leg of the trapezoid can accommodate the sharp angle created by the edges of the 3D grid. An interesting extension is to build other regular meshes using a similar approach \cite{akleman2006regular,akleman2005regular,van2009symmetric}. The Connectors and the corresponding Edge Elements for these will also be one type. Therefore, they can also be mass-produced. 
\item \textbf{Trusses based on Arbitrary Polygonal Meshes:} For meshes that represented arbitrarily shaped surfaces, the design of the Edge Element would naturally be dictated by the curvature characteristics of the surface. Here, we envision that the curvature of such shapes can be achieved by altering the cross-section of the toroidal domain based on the incoming edge. This leads to the creation of a torus with a unique cross-section at every point along its toroidal direction. The sinusoidal curves also need to be organized to fit all the incoming edges at a vertex. However, there is a need for solving both theoretical and practical problems starting from reducing the number of the different elements for economical manufacturing. 
\item \textbf{Mechanical Characterization:} We believe that our Edge Elements, while primarily envisaged for artistic applications, could make a potentially powerful candidate for engineering applications. Indeed, our inspiration to create trusses came directly from structural mechanics. Having said that, a series of deeper numerical, as well as physical experiments are necessary to understand the fundamental behavior of our Connectors and Edge Elements. This could even prove useful in the very design of our proposed structures by allowing us to strategically distribute the key-in-hole locations on the Connectors in an effective way.
\end{itemize}

\bibliographystyle{abbrv}
\bibliography{references, refarch, wovenTiles}
\end{document}